# Chatbot for fitness management using IBM Watson

Sai Rugved Lola, Rahul Dhadvai, Wei Wang, Ting Zhu
University of Maryland Baltimore County
an63211@umbc.edu  rahuld1@umbc.edu  ax29092@umbc.edu  zt@umbc.edu

***Abstract—*** *Chatbots have revolutionized the way humans interact with computer systems and they have substituted the use of service agents, call center representatives etc. Fitness industry has always been a growing industry although it has not adapted to the latest technologies like AI, ML and cloud computing. In this paper, we propose an idea to develop a chatbot for fitness management using IBM Watson and integrate it with a web application. We proposed using Natural Language Processing (NLP) and Natural Language Understanding (NLU) along with frameworks of IBM Cloud Watson provided for the Chatbot Assistant. This software uses a serverless architecture to combine the services of a professional by offering diet plans, home exercises, interactive counseling sessions, fitness recommendations.*

***Keywords—****NLP, Intent Classification, Entity Recognition, Personalization, Human Computer Interaction, Conversational technology.*

## I. INTRODUCTION

Businesses benefit from chatbots because they increase operational efficiency and save costs while providing convenience and extra services to internal staff and external clients. They enable businesses to quickly answer a variety of client concerns and enquiries while decreasing the requirement for human involvement.

A firm can expand, customize, and be proactive all at the same time using chatbots, which is a key differentiation. When a firm relies exclusively on human power, for example, it can only service a certain number of people at any given time. Human-powered firms are compelled to rely on standardized models in order to be cost-effective, and their proactive and personalized outreach skills are restricted.

Chatbots, on the other hand, allow businesses to interact with an endless number of consumers in a personalized manner and can be scaled up or down based on demand and business needs. A company may give humanlike, tailored, proactive service to millions of customers at the same time by deploying chatbots.

Our focus in this paper is to implement an interactive chatbot for personal fitness management. To assist users in yoga plan, diet plan, scheduling, users can use this interactive and advanced chatbot. The origins of the chatbot may be traced back to Alan Turing's idea of sentient robots in the 1950s. Since then, artificial intelligence, which is the foundation for chatbots, has advanced to encompass super intelligent supercomputers like IBM Watson.

Chatbots are widely used to improve the IT service management experience, which focuses on self-service and automating internal operations. Common activities such as password changes, system status, outage notifications, and knowledge management may be easily automated and made available 24 hours a day, seven days a week with an intelligent chatbot, while expanding access to widely used voice and text-based conversational interfaces.

Chatbots are most typically employed in customer contact centers to manage incoming conversations and guide consumers to the right resource. Internally, they're typically used for onboarding new workers and assisting all staff with routine tasks like vacation scheduling, training, purchasing computers and business supplies, and other self-service activities that don't require human participation.

The possibilities they open up in the fitness business are practically limitless, but for now, they're mostly being used to meet initial marketing and communication objectives.

The rest of the paper is organized as follows. Section-II describes the problem statement, in section-III we also identified the related work on the proposed work, Section-IV we explained the implementation of the project and internals of the IBM Watson, in Section-V we proposed the future works that are to be implemented, Section-VI we described the conclusion and Section- VII are the References.

## II. PROBLEM

Although going to the gym has several advantages, it is not for everyone. Approximately 67 percent of gym members do not go on a regular basis. People are apprehensive about unfamiliar environments or using difficult-to-use fitness devices if they haven't been properly trained. Introduction sessions are available at certain gyms, whereas personal trainers handle this at others.

In addition, there is a lack of interaction between gyms and gym members. Gym members are not incented or reminded to go to the gym on a regular basis in any meaningful way. Filling up the gym's facilities is really against the gym's interests.

Unfortunately, few gyms are built with a holistic health approach in mind. When emotional and mental health aren't supported jointly, it's challenging to achieve peak physical performance. Reduced physical performance results in poor outcomes, making it difficult to establish a regular gym program.

There are several fitness applications available that focus on delivering performance feedback, food recommendations, and activity routines. Fitness applications can be beneficial to those who are willing to put in the extra effort to use the program. Unfortunately, few people have the mental capacity to stay engaged with applications and enter personal fitness data, diets, or training regimens on a daily basis. Although analytics might aid in personal performance optimization, some users are intimidated by the prospect of monitoring statistics too frequently. Daily body weight swings, as well as failure to improve on prior weeks' performance, might sap motivation.

We aim to provide a technological solution to people who feel that maintaining a regular diet plan or a workout routine is tough and also an overhead to remember it all the time. Instead, users can use our chatbot to schedule all their workout plans and diet plans which will be saved in the database.

Education, constant reminders, compelling material, and prizes all help to motivate people to work out more often. These features add to the user's mental well-being and aid in the development of good habits. A constant accountability partner who reminds the user of their own objective increase's motivation significantly. We hope to mimic an accountability partner that sends out interesting push alerts on certain days by constructing a chatbot. A personal fitness trainer is a person who helps people become in shape.

Self-efficacy may be adjusted and strengthened with the correct stimulus, according to psychological research (Weinberg et al.,1979).

Finally, users must be rewarded for their efforts in order for their input to be linked to a beneficial outcome. While we're working on the feature roadmap, a rewards function and additional material will be added to the chatbot shortly.

The aim is to provide a tailored chatbot experience that isn't time consuming and keeps the customer engaged in their weekly exercise routines.

## III. RELATED WORK

The advanced techniques have enabled smart applications in different scenarios [15-]. According to the International Health, Racquet &

Sports club Association, the worldwide fitness sector generated $83.1 billion in sales in 2016, up 2.6 percent from $81 billion the previous year. It should reach $87.5 billion in 2018 if it continues at this pace. Currently, a number of AI companies claim to be able to help gyms with new client acquisition, nutrition and fitness program development, and customer relationship management. Some companies believe that chatbots can assist gym members in sticking to their own fitness plans.

Vida Health says that their software can assist people and business personnel maintain their health and fitness by connecting them with a one-on-one coach using a machine learning-based app.

Vida Health says that their app can connect people with a Vida Health coach or therapist based on their health objectives, preferred coaching style, biometrics, availability, and geographic region.

Users must first input information about their health objectives, whether they are connected to physical condition, weight reduction, or stress alleviation, before they can begin using the program on their mobile phone. They must also pick a coaching style that suits them best. Based on the data, the app's machine learning model searches its database for traits that are similar to those picked by the user and matches them with a coach who can assist them in achieving their objectives.

The business says that its methodology can assist trainers communicate with clients by providing push alerts and remotely tracking client progress via wearables and smart monitor integrations.

Gym management, for example, will be able to see previous years' bookings for class spots. Managers may estimate which classes and time slots will be popular in the future using this information. The algorithm will be able to forecast a member's success in meeting their goals by scanning its database for outcomes with comparable qualities by combining data logs on nutrition, workout performance, age, weight, and gender. It also directs trainers to suggest additional exercises or dietary regimens if necessary.

Zara is a smartphone application developed by Zova, an Australian business that claims to give fitness fans with smart, tailored exercises via a chatbot. Zara tailors training routines to specific consumers, according to the firm.

The app also has an Activity Score function that calculates an intensity level for these activities by averaging the user's activity level over seven days, including all steps and exercises.

The algorithms may then search their database of trainings and workouts for exercises that are comparable to the user's lifestyle and suggest workouts that will assist the user in meeting their fitness objectives. Zova may be used to webcast and track the workouts.

Fitwell is a firm headquartered in the United Kingdom that promises to be able to assist people measure their fitness progress with a chatbot called Hailee.

The software includes training services tailored to the users' specific goals, such as becoming stronger, more flexible, or toned. The algorithms in the app will go through its database for training programs that fit the user's preferences. The algorithms will alter the exercises over time based on the user's prior performance to ensure that they meet their objectives.

The software also features a bot that, according to the business, allows users to keep track of their meals and snacks by delivering accurate nutritional information.

## IV. IMPLEMENTATION

Conversational assistant aids in the automation of chat dialogues between users and computers. This chatbot is meant to address fitness needs and assist individuals in obtaining information about trainers and diet plans from the comfort of their own homes. These interactive chatbots will aid in the resolution of difficulties such as the lack of information on the proper food plan, exercise comprehension, and body care.

### A. Chat Bot Architecture

In recent years, serverless architectures (Functions-as-a-Service) have gained traction as a way to deliver backend services without requiring a dedicated infrastructure. Serverless allows users to integrate stateless functionality into platform systems. Each invocation is independent of the previous ones due to its statelessness. IBM Watson cloud functions has been used as the backend infrastructure for our application.

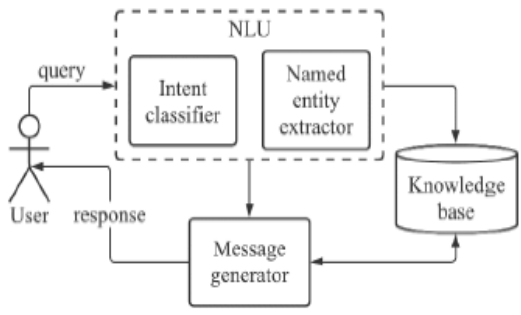

Fig 1: Simple Chatbot Components

Our conversational bot is AI-powered and developed with serverless architecture. It has Natural Language Processing (NLP) and Natural Language Understanding (NLU) [12] built in to comprehend the user's inquiry and respond appropriately. NLP makes it easier to read, decode, comprehend, and interpret human languages. To interpret the semantic meaning of user input and locate morphemes, the user query is processed and decoded. The intents are detected and then mapped on to the Dialog skill in IBM Watson.

The intent mapping is then handled by the Dialogskill [4] which has been thoroughly trained with the different dataset and questionnaires and helps in drawing the inferences. The Watson Assistant V2 API are used to generate a HTTP POST request in order to send the query from the application to server. The request consists of query dialogskill then decodes the intent to understand actions, entities and then maps it to intent to define the response. The Dialogskill then sends the response back for the request with the help. The response message contains the response that the user should receive at the conclusion of the operation.

The main components of IBM Watson includes the following

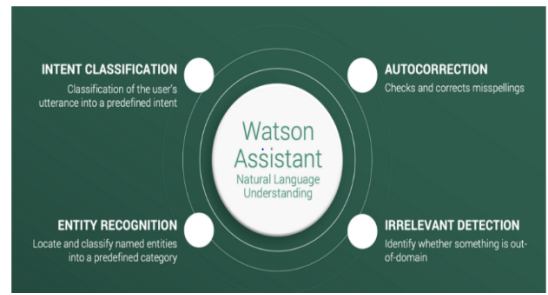

Fig 2: Components of IBM Watson.

### B. Intent Classification:

A purpose or objective conveyed in a customer's input, such as answering a question or pushing a bill payment button, is known as intent. The Assistance service selects the appropriate dialog flow for responding to a customer's input by recognizing the intent indicated in it. It's a technique for steering the conversation in the right direction.

[8] The query formulated by the user must be analyzed by the intent classifier and must be able to distinguish whether it is in-scope (IS) or Out-Of-Scope (OOS), NLP plays a major role in classification of query correctly. Here we also have the concept if reformulation which will be discussed further.

The Dialogskill first does matching using the Rule -Based grammar matching, if no match found switches to the ML algorithmic approach, like Support vector Machine for understanding the semantics in the query formulated by the user.

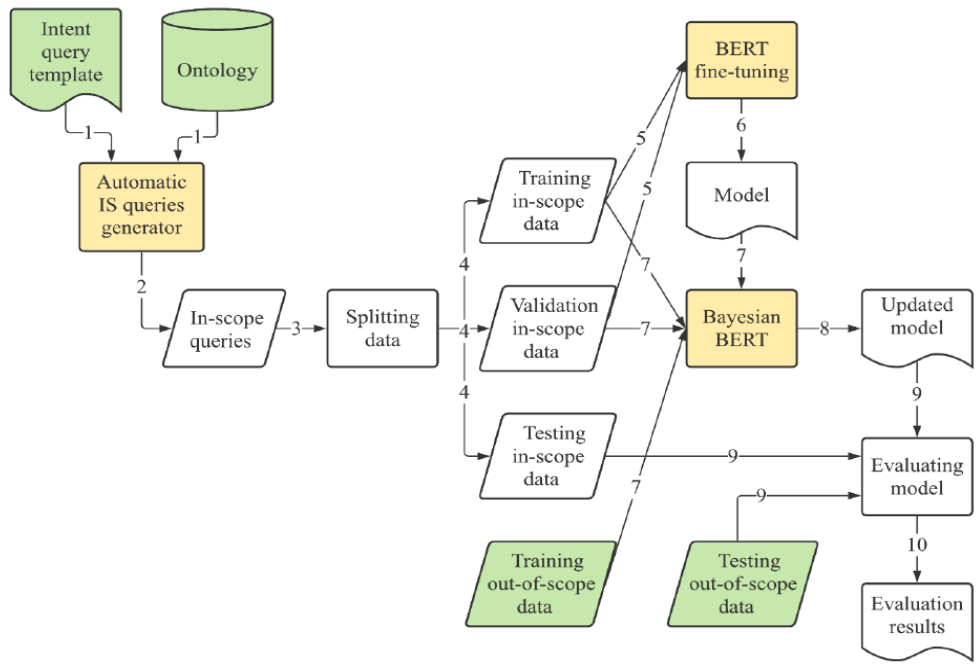

Fig 3 [8]: Procedure for Intent detection.

The AutoML uses the set of algorithms like Bidirectional Encode Representation from Transformers (BERT) [8] for the purpose of NLP pretraining. The Transformer is a focus mechanism that learns how words in a text are related to one another. It was trained unsupervised using a plain text library, such as Wikipedia or other freely available text data on the internet in any language. As a result, in any NLP classification challenge, BERT outperforms traditional ML systems. Here we use the confidence score in order to understand the relevancy of query with the intent, ranging from 0.0 (completely uncertain) and 1.0 (fully relevant). If the confidence score of the highest scoring IS intent is greater than or equal to the threshold setting, it is returned to match. Otherwise, an OOS is matched if no IS intents meet the criteria.

The intent detection also uses Meta-Learning [11] algorithms for Few-Shot Computer Vision to train from the larger dataset, and learn from them and form some predefined contextual intent for knowledge base and use this to work on newer, unseen queries.

### C. Irrelevance Detection Algorithm

Another difficult difficulty in conversational AI is determining when end users are asking about off-topic topics. A banking virtual assistant, for example, should not be expected to respond to inquiries about dinosaurs and should gently divert the user back to banking.

Watson Assistant, as we discussed in the intent detection section above, requires relatively minimal training data to effectively grasp user intents. However, recognizing in-domain and out-of-domain borders becomes more difficult as a result.

To accomplish so, we use deep learning models to model the training data's distribution by translating it to a domain representation. Any input whose representation does not match the distribution is considered irrelevant based on the learnt model. This implies that our algorithms are well-versed in the types of sentences users are likely to employ while discussing banking. When the models come across terms that aren't commonly used in banking, they recognize that the discourse isn't on track.

*D. Autocorrection Algorithm*

Remember, utterances are the many ways in which end users might ask a specific inquiry or request a certain action — for example, "password doesn't work" or "I can't log in" are two utterances that map to the "reset password" intent.

Because so much depends on accurately comprehending utterances, an autocorrection method is required to repair misspelled words. It can't be harsh, though, because business jargon frequently uses abbreviations and trademarks that are spoofs of actual words. Profanity screening, for example, is also required. All of this, and more, is available through Watson Assistant.

Watson Assistant is driven by language models, phonetic models, edit distance models, and deep learning under the hood, allowing it to correct a wide range of misspellings in corporate use cases.

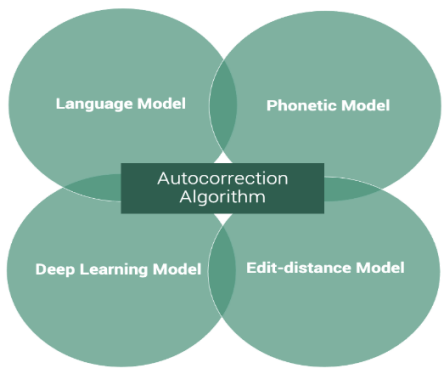

Fig 4 [7]: Autocorrection Algorithm components

*E. Entity Recognition Algorithms*

In the framework of conversational AI, let us define "entity." "IBM Watson Discovery" is a software product that we want the virtual assistant to pay special attention to. We want the system to comprehend "I need help with discovery" in the context of a product problem chatbot.

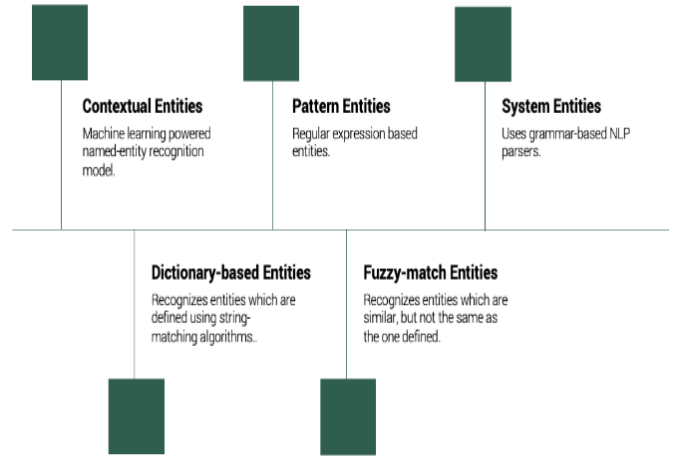

Fig 5 [7]: Types of Entity Recognition.

In Watson Assistant, we have a variety of entity recognition methods.
- Contextual entities allow your virtual assistant to recognize entities depending on the context of the user's speech by applying deep learning features to train a named-entity recognition model.
- Using grammar-based NLP approaches, system entity algorithms recognize date, time, range, number, currency, and other data in user utterances.
- The administrator defines particular names, synonyms, or patterns for dictionary-based entities. Only when a phrase in the user input perfectly matches the value or one of its synonyms does the assistant identify the entity mentioned at run time.
- Fuzzy match entities identify phrases that have similar syntax to the entity value and synonyms you supply, but don't need a precise match.
- The user can construct a regular expression to match the entity when using a pattern-based entity.

Building these algorithms is a difficult process, especially when size is factored in. Furthermore, product limits to make it more user-friendly — such as allowing for shorter training cycles to make it seem more responsive — contribute to the difficulty of creating and building NLP algorithms.

By reformulating initial user inquiries based on the user context and profile, we hope to increase the precision of information retrieval systems.

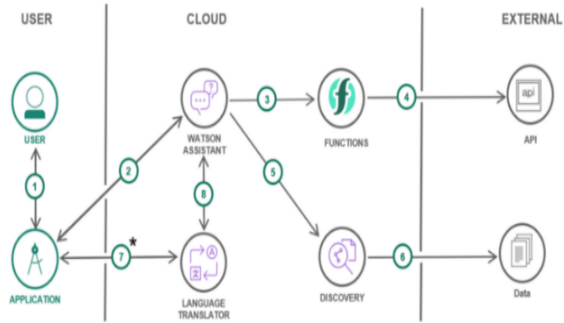

Fig 6 [10]: Application Architectural Flow

### F. User's Profile and Context

When a user conducts a search, identifying and describing his or her working context may be reduced to identifying his or her current task and identifying associated terms from his or her profile. This is based on observing the current user's task as a contextual component (for example, searching for a restaurant or hotel, planning a trip, and so on). As a result, we create an intelligent assistant that extracts relevant phrases from the current search session, but what exactly do we mean by relevant terms?
If a term is comprehensive and precise, it is relevant

• Complete: The keywords are linked to a submitted query, user profile, and task at the same time. (Expanding the query.)
• Exact: no stop words, repeated terms, or out-of-context terms are included in the terms.

These phrases are utilized to create a new reformulated query that will be sent to the information retrieval system in order to deliver context-based results. These phrases are not required to be associated with the next search session at the same user's task.

Here, we'll go over our method, which consists of three models: a task model, a user profile model, and a State Reformulated Queries (SRQ) model.

The task model is in charge of creating the current working context by selecting one task from a list of preset tasks to attach to the first query. The user profile model is in charge of utilizing the user profile by adapting the obtained results to this user using information included in the profile.
The SRQ model is in charge of gathering characteristics from the current task, one attribute for each task state at the very least. The operational profile may be used to get the values of these attributes. To reformulate a user query, we first expand the query with relevant phrases before removing the unnecessary terms (query refinement). SRQ stands for the resultant query (State reformulated Query).

### G. User Context Modeling

In this part, we'll provide a novel contextual analysis approach that considers the user's present task and how it evolves over time as the user's context. The shift from one stage to the next indicates that the user has finished this step of the current job. Thus, when we discuss the user context in this study, we are referring to the task that the user is working on at the time the information retrieval process happens, as well as the stages of that job. As a result, we must model the user's present task in order to broaden the user query with contextual task phrases that direct the search to the most relevant results.

As one of the contextual aspects that surround the user during the information retrieval process, the task model is used to detect and define the task that is done by the user when he sends his or her query to the information retrieval system.

To begin, we must separate the action from the job. In actuality, an activity might be something you're merely doing, with or without a purpose; it's the action you're taking, but a task is the goal you've set for yourself. In order to complete the mission, the activities are necessary. To put it another way, a task is a set of tasks that may or may not be completed. As a result, we may use a UML activity diagram to describe the user's job, which incorporates all of the actions required to complete it. The task state refers to the several stages required to complete the present work.

We rely on research questionnaires [5] in our task model, which were intended to generate tasks that were predicted to be of interest to individuals throughout the study. Subjects were instructed to think about their online information seeking behaviors in terms of tasks and to establish personal labels for each task in that research [5]. They were given some examples of activities to do, such as "writing a research paper," "traveling," and "shopping," but they were not directed, influenced, or prejudiced in their task selection in any way. The following nine task categories were created using a general classification system for all tasks specified by all subjects:

### H. Conversational Flow:

Our bot's communication has been framed and intended to emulate human behavior in order to provide a user-friendly chat system that puts people at ease while overcoming the bias of machine contact.

Countless utterances that users can use while communicating with the chat system are taken into account while building conversational processes.

The Dialogskill Discussion API[14] is used in our application to create an automated conversational chat system that understands natural language and holds a conversation with the user. Dialogskill makes use of IBM Cloud Machine Learning technology, which enables apps to scale to hundreds of millions of users. When a user communicates with our software, they can use a wide range of expressions to achieve the same goal. Dialogskill, in example, offers a high-level dialogue flow for identifying user queries by mapping them to intents that have been thoroughly trained.

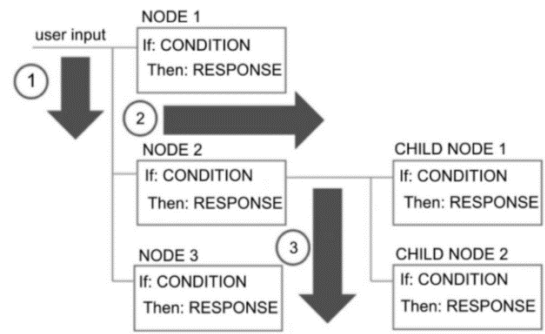

Fig 7 [10]: Conversational Flow

### I. Issues Encountered

The challenges faced during the development of application includes a) Collection of data for training. The regular machine learning applications are trained using large amount of data sets which will cover most of the scenarios while the chatbots are trained on the conversations, intent detection systems are practically relied on the conversational data from the customers. So, collection of relevant data and training the model on the conversational basis is a challenging task to be done.

b) Using different phrases and jargons by different individual to address the same problem or query. When used in real-world, chatbots They frequently work with test data that is drastically different from the training data. The difference in train and test data distributions is mostly due to the free-form nature of input user queries. Because these real-world questions use non-standard paraphrases to describe the same goals, it's impossible to cover everything during training. The problem is exacerbated by the unavailability of large, high-quality training data

C) Efficiency in Computation, the requirements from the user changes every moment and the retraining from the conversations and analyzing the new queries and inferencing and responding must be quick. So efficiency in responding and with contextual accuracy is the most challenging task.

### V. FUTURE WORKS

*A. Integrating Chatbots with Applications*

Chatbots helps in interacting the user with the application in the swiftest way and they also make the application most interactive. These are helpful in maintaining a healthy relationship between the customer and application. Chatbots can be integrated in different applications like delivery, telecommunications, health care, web applications. They can also be integrated with applications like calendar, reminder and fitness tracking. By this, the users can easily communicate with the device regarding their appointments, scheduled meetings, and also to evaluate their fitness goals. Example is BookMyshow integrating a normal chatbot with WhatsApp to deliver the tickets online. Likewise, we can use Chatbot for fitness applications to remind about the exercise, diet plans to be followed and guide them. With the help of services like Amazon Web Services, Google services, IBM Watson the integration, scalability and maintenance of the chatbots have become easier.

*B. Conversation-Driven Recommendation*

Chatbots can also be the source for information about the customer interests. The information collected from user can be grouped together and the data can be used to understand about the trend and recommend the necessary suggestions for the user. The chatbot can recommend the diet plans, exercises by understanding the conversational flow and by learning from the previously asked questions using the Natural Language Processing (NLP) and building the relationship between user and the intent. The process has a problem "cold start problem" which means the process needs some data about customer to recommend. In order to overcome this problem, we recommend the most followed plans on which algorithm is trained. So, this would involve in the granular personalization at user level.

*C. Scalable Automation with Multi-Part Queries*

The Chatbots are replacing the Customer Service Individuals in every sector, so scalability and also understanding the query of the user is important to respond accurately and with speed. This study may be expanded in a number of ways. First, improve the quality of produced words, then the precision of outcomes; second, improve the task model to improve the identification of the user's task and its states. We may leverage the Chatbot technique to improve the assistant's capacity to create reformulated inquiries (SRQ), which allows the assistant to engage with a user while maintaining focused on the present task condition. We also plan to put this method to the test using a different evaluation protocol that entails creating a test collection and determining relevant results for several queries in a specific context, then comparing these relevant results to the results returned by our system for the same queries in the same context.

*D. Multilingual Chatbots*

To guarantee that a firm achieves its full potential, impediments such as language issues must be removed from the client journey. The impact on a firm will be disastrous if potential leads have communication challenges or if an inadequate solution is supplied simply because the user requirements were misunderstood. Language detection can be identified by using the Named Entity Detection by using the NLP algorithms, changing the SDK's supporting libraries for each language and training the chatbot in different languages rigorously will help in improving the conversations and reduce the error rate or non-contextual replies.

*E. Voice Bots*

Voice Bots make the communication between the customer and application touch-free making everything virtualized. Voice bots are artificial intelligence (AI)-powered software that allows a caller to navigate an interactive voice response (IVR) system using only their voice and natural language. Callers are not required to listen to menus or input digits on their keypads. They communicate with the IVR in a simplified simulation of a live operator. Voicebots go beyond limiting short-term remedies like

Automated Speech Recognition (ASR) with predetermined expressions to improve voice interaction between callers and IVR. The process involves speech recognition and then processing the data using Natural Language Processing (NLP) and then training these bots on analysis, translation. These will improve the accuracy rate, customer quick and direct interactions.

## VI. CONCLUSION

Artificial intelligence (AI) and machine learning have revolutionized the way we do business in recent years. The market has been captured by chatbots. Chatbots are now commonplace. Financial Services, Retail, Fitness Management, hospitals have all had breakthroughs as a result of its business implications. Chatbots can now manage client interactions 24 hours a day, seven days a week, while continually increasing the quality of replies and lowering expenses. Chatbots automate processes and relieve staff of tedious chores. It involves collection of relevant data for contextual accuracy, Using Deep Learning frameworks, NLP algorithms like "Multinominal Naïve Bayes" in order to train the chatbot on dataset on understanding users sentimental analysis, Intent detection and automate the process in order to improve the contextual accuracy. re-learning is most important aspect for improving the chatbot user experience. In this way, chatbots will build customer relationships, create brand awareness and automates tedious work and plays a huge role in shaping the business.